\documentclass[journal,twocolumn,letterpaper]{IEEEtran}
\pdfoutput=1
\usepackage{amssymb}
\usepackage{amsmath}
\usepackage{graphicx}
\usepackage[amsmath,thmmarks]{ntheorem}

\newcommand{\C}{\ensuremath{\mathcal{C}}}
\newcommand{\G}{\ensuremath{\mathcal{G}}}
\newcommand{\A}{\ensuremath{\mathcal{A}}}
\newcommand{\B}{\ensuremath{\mathcal{B}}}

\newcommand{\calS}{\ensuremath{\mathcal{S}}}

\newcommand{\T}{\ensuremath{\mathcal{T}}}

\newcommand{\rank}{\mathrm{rank}}

\newcommand{\comment}[1]{}
\newcommand{\ceil}[1]{\left\lceil #1 \right\rceil}

\newtheorem{theorem}{Theorem}
\newtheorem{lemma}{Lemma}
\newtheorem{corollary}{Corollary}
\newtheorem{proposition}{Proposition}
\theorembodyfont{\upshape}
\newcounter{mycnt}

\newtheorem{construction}[mycnt]{Construction}
\newtheorem{definition}{Definition}

\allowdisplaybreaks[1]

\def\mytitle{Single-Exclusion Number and\\
the Stopping Redundancy of MDS Codes}
\def\figscale{1}

\begin{document}

\title{\mytitle}
\author{Junsheng~Han~\IEEEmembership{Student~Member,~IEEE},%
        ~Paul~H.~Siegel,~\IEEEmembership{Fellow,~IEEE},%
        ~and~Ron~M.~Roth,~\IEEEmembership{Fellow,~IEEE}%
\thanks{J.~Han and P.~H.~Siegel are with the University of California, San Diego, 
        La Jolla, CA 92093-0407 (e-mail: han@cts.ucsd.edu, psiegel@ucsd.edu).}%
\thanks{R. M. Roth is with the Computer Science Department, Technion--Israel
Institute of Technology, Technion City, Haifa 32000, Israel (e-mail: ronny@cs.technion.ac.il).}}
\maketitle

\begin{abstract}
For a linear block code $\C$, its \emph{stopping redundancy} is defined
as the smallest number of check nodes in a Tanner graph for $\C$,
such that there exist no stopping sets of size smaller than the minimum distance of $\C$.
Schwartz and Vardy conjectured that the stopping redundancy of an MDS
code should only depend on its length and minimum distance.

We define the \emph{$(n,t)$-single-exclusion number, $S(n,t)$} as the smallest
number of $t$-subsets of an $n$-set, such that for each $i$-subset of 
the $n$-set, $i=1,\ldots,t+1$, there exists a $t$-subset that contains all but
one element of the $i$-subset.
New upper bounds on the single-exclusion number are obtained via probabilistic
methods, recurrent inequalities, as well as explicit constructions.
The new bounds are used to better understand the stopping redundancy
of MDS codes.
In particular, it is shown that for $[n,k=n-d+1,d]$ MDS codes,
as $n\rightarrow\infty$, the stopping redundancy is asymptotic to $S(n,d-2)$,
if $d=o(\sqrt{n})$, or if $k=o(\sqrt{n})$, $k\rightarrow\infty$, thus giving partial confirmation of the Schwartz-Vardy conjecture in the asymptotic sense.
\end{abstract}


\IEEEpeerreviewmaketitle

\section{Introduction}

The stopping redundancy of a linear code characterizes the minimum ``complexity'' (number of check nodes) required in a Tanner graph for the code, 
such that iterative erasure decoding achieves performance comparable to 
(up to a constant factor, asymptotically) maximum-likelihood (ML) decoding.
It can be viewed as a basic measure of the complexity-performance tradeoff in the use of redundant parity checks (RPCs) in an iterative decoder on the erasure channel.

Although this tradeoff is less straightforward to understand in non-erasure channels, there is empirical evidence that RPCs can improve performance in belief-propagation decoding on an AWGN channel \cite{stop:kou01}, \cite{stop:kelley07}, and, recently, the concept of stopping redundancy has provided partial motivation for novel decoding algorithms that achieve near-ML performance for short, high-rate codes \cite{stop:hehn07b}. 

Formally, we define stopping redundancy as follows.
Let $\C$ be an $[n,k,d]$ linear code,
and let $H=(h_{ij})_{l\times n}$ be a parity-check matrix for $\C$.
We shall assume that $\rank(H)=n-k$, but $l$ may be larger than $(n-k)$.
The Tanner graph $\G(H)$ is a bipartite graph
with $n$ variable nodes, each corresponding to one column of $H$,
and $l$ check nodes, each corresponding to one row of $H$, 
such that variable node $j$ is adjacent to check node $i$ if and only if
$h_{ij}\neq 0$.
A \emph{stopping set} in $\G(H)$ is a set of variable nodes such that 
all check nodes adjacent to the set are connected to the set at least twice.
It is well known \cite{stop:di02} that iterative erasure decoding is successful
if and only if the set of erasure locations does not contain a stopping set.
The size of a smallest non-empty stopping set, referred to as 
the \emph{stopping distance} and denoted by $s(H)$, is therefore an important parameter governing the performance of the iterative decoder.
It is clear that $s(H)\leq d$, and it not difficult to see that equality can
be achieved for any code, for example by choosing the rows of $H$ to be the non-zero codewords of the dual code $\C^{\perp}$.  This leads to the following definition.

\begin{definition}
Let $\C$ be a linear code with minimum distance $d$.
The \emph{stopping redundancy} of $\C$, denoted by $\rho(\C)$,
is the smallest integer such that there exists a parity-check matrix $H$
for $\C$ with $\rho(\C)$ rows, and $s(H)=d$.
\end{definition}

Stopping redundancy was introduced by Schwartz and Vardy
\cite{stop:schwartz05}, \cite{stop:schwartz06}, and was further studied
in \cite{stop:etzion06}, \cite{stop:han07a}.
The concept was later extended in a number of interesting ways
\cite{stop:milenkovic06}, \cite{stop:hehn06}, \cite{stop:ghaffar07}.
Related concepts, such as \emph{stopping set enumerator}, 
and \emph{generic erasure-correcting sets}, were studied in
\cite{stop:weber05}, \cite{stop:weber06s1},
and in \cite{stop:hollmann06}, \cite{stop:hollmann07}, respectively.

In this paper, we focus on the special class of MDS codes.
In the rest of the paper, unless otherwise noted, 
$\C$ denotes an $[n,k=n-d+1,d]$ linear MDS code.
In \cite{stop:schwartz06}, it was shown that for all $d\geq 3$,
\begin{equation}\label{eqn0}
\frac{1}{d-1}{n\choose d-2}
< \rho(\C)
\leq \frac{\max\{d^\perp,d-1\}}{n} {n\choose d-2},
\end{equation}
where
\[
d^\perp = n-d+2
\]
is the minimum distance of $\C^\perp$, the dual code of $\C$.
The authors of \cite{stop:schwartz06} then made an intriguing
conjecture that $\rho(\C)$ should in fact be just a function of $n$ and $d$.

Note that the upper bound in \eqref{eqn0} is never better than
\[
\frac{1}{2} {n\choose d-2}.
\]
So the upper and lower bounds can differ by
up to a factor of $n$.
In \cite{stop:han07a}, we observed that the upper bound can be improved
by introducing a new combinatorial quantity, the single-exclusion number,
which we describe below. 
Before doing so, we first review two related, well-studied combinatorial constructs.
For positive integers $n\geq s\geq t$, and an $n$-set%
\footnote{An \emph{$n$-set} is a set that contains $n$ elements.
Similarly, if $A$ is any set, then a \emph{$t$-subset} of $A$ is a subset of $A$
that contains $t$ elements.}%
$N$, an \emph{$(n,s,t)$-Tur\'an system} \cite{stop:sidorenko97}
is a collection of $t$-subsets of $N$, called \emph{blocks}, 
such that each $s$-subset of the $n$-set contains at least one block.
The \emph{$(n,s,t)$-Tur\'an number}, denoted hereafter by $T(n,s,t)$,
is the smallest number of blocks in an $(n,s,t)$-Tur\'an system.
A concept ``dual'' to that of a Tur\'an system is that of a covering design \cite{stop:mills92}.
Specifically, for $n\geq s\geq t$ and an $n$-set $N$, 
an \emph{$(n,s,t)$-covering design} is a collection of $s$-subsets of $N$,
also called \emph{blocks}, such that each $t$-subset of the $n$-set
is contained in at least one block.
The \emph{$(n,s,t)$-covering number}, denoted by $C(n,s,t)$,
is the smallest number of blocks in an $(n,s,t)$-covering design.
Clearly, by definition,
\[
T(n,s,t)=C(n,n-t,n-s).
\]

The stopping redundancy of an MDS code is closely related to
covering/Tur\'an numbers.
In fact, the lower bound in \eqref{eqn0} was shown by noting that
to maximize $s(H)$, the supports of minimum-weight rows of $H$
must form an $(n,n-d+2,n-d+1)$-covering design (equivalently, the 
complements of supports form an $(n,d-1,d-2)$-Tur\'an system).
Hence,
\[
\rho(\C)\geq C(n,n-d+2,n-d+1)=T(n,d-1,d-2).
\]

We now define the single-exclusion number,
which was introduced in \cite{stop:han06}.

\begin{definition}
For an $n$-set $N$ and $t<n$, 
an \emph{$(n,t)$-single-exclusion (SE) system} is a collection of $t$-subsets of $N$,
called \emph{blocks}, such that for each $i$-subset of $N$, $i=1,\ldots,t+1$,
there exists at least one block that contains all but one element from the $i$-subset.
The \emph{$(n,t)$-single-exclusion (SE) number}, $S(n,t)$, is the smallest number of blocks in an $(n,t)$-SE system.
\end{definition}

Note that an $(n,t)$-SE system is a special kind of $(n,t+1,t)$-Tur\'an system.
Note also that if we require all rows in the parity-check matrix
$H$ to be of minimum weight, then $s(H)=d$ is equivalent to the condition  
that the sets of column indices corresponding to
the zeros in each row form an $(n,d-2)$-SE system.
Hence,
\[
\rho(\C)\leq S(n,d-2).
\]
Therefore, any upper bound on $S(n,d-2)$ is an upper bound on $\rho(\C)$.

In \cite{stop:han07a}, a number of upper bounds on $S(n,t)$ were obtained
using combinatorial constructions and were shown to be superior to the upper bound
in \eqref{eqn0}.
It was shown that $\rho(\C)=T(n,d-1,d-2)$ for $1<d\leq 4$, $n\geq 6$, 
and $\rho(\C)\leq T(n,d-1,d-2)+1$ for $d=5$.
It was also shown that $\rho(\C)$ is asymptotic to $T(n,d-1,d-2)$ 
(and to $S(n,d-2)$) as $n\rightarrow\infty$ for any fixed $d$,
and that it is asymptotically at most $3T(n,d-1,d-2)$ for any fixed $k=n-d+1$.

In this paper, we build upon the work in \cite{stop:han07a} and investigate
$S(n,t)$ through a number of different approaches. 
New upper bounds are obtained and analyzed.
They are then used to show that as $n\rightarrow\infty$,
$\rho(\C)$ is asymptotic to $S(n,d-2)$ if $d=o(\sqrt{n})$,
or if $k=o(\sqrt{n})\rightarrow\infty$.%
\footnote{
We adopt the standard ``O-notation'' and related asymptotic expressions
\cite[Ch. 9]{stop:graham94}.
Functions $f(n)$ and $g(n)$ are said to be \emph{asymptotic}
to each other, denoted by $f(n)\sim g(n)$,
if $\lim_{n\rightarrow\infty} f(n)/g(n)=1$,
or equivalently, if $f(n)=\bigl(1+o(1)\bigr)g(n)$,
where $o(1)$ stands for any function that goes to zero as $n$ goes to infinity.
More generally, we write $f(n)=o\bigl(g(n)\bigr)$
if $\lim_{n\rightarrow\infty} f(n)/g(n)=0$.
}
Hence, in an asymptotic sense, the Schwartz-Vardy conjecture is proved in these cases.
For all $d\leq 5$, it is shown that $\rho(\C)=S(n,d-2)$.
A lower bound on $S(n,t)$ is also derived.

Besides their application to the stopping redundancy of MDS codes, SE systems warrant further study for other reasons. 
As combinatorial objects, they have a very natural definition. 
Therefore, they have intrinsic mathematical appeal and interest. They also have practical relevance; for example, the definition of SE system can be readily mapped to a problem in the design of experiments.

We remark, also, that some of the results on SE numbers are rather surprising.
For example, we shall see that for all $k=o(\sqrt{n})$,
\[
\frac{S(n,n-k-1)}{T(n,n-k,n-k-1)} \leq 
\bigl(1+o(1)\bigr) \frac{k+1}{k}.
\]
Thus, despite the apparently much stricter requirements imposed
upon the SE system in comparison to the Tur\'an system, the increase
in the total number of blocks is very small.
Insights such as this may shed further light upon properties of both MDS
codes and the various combinatorial constructs to which they are closely related.

The rest of the paper is arranged as follows.
Section~\ref{sec1} is devoted to upper bounds on $S(n,t)$.
Three approaches are attempted: combinatorial constructions, 
probabilistic methods, and recurrent inequalities.
Asymptotics, especially those based on results from recurrent inequalities,
are discussed.
Explicit (calculable) upper bounds are compared, and the best bounds
are found for $n$ up to $512$.
In Section~\ref{sec2}, we show a lower bound on $S(n,t)$.
In Section~\ref{sec3}, we comment on the Schwartz-Vardy conjecture.
We summarize what is known about the conjecture based on the results
in this paper and previous works, and comment on our conjecture
made in \cite{stop:han06} that $\rho(\C)=S(n,d-2)$.
Section~\ref{sec4} concludes the paper.

\section{Upper Bounds on $S(n,t)$}\label{sec1}

We start with some preliminaries.

For any set $A$, let $[A]^i$ denote the set of all $i$-subsets of $A$.
If $A$ and $B$ are sets, we say that $A$ 
\emph{covers} $B$ if $|B\setminus A|=1$. 
Hence, if $N$ is an $n$-set, then $\calS\subseteq [N]^t$ is an $(n,t)$-SE system
if and only if for each $i=1,\ldots,t+1$ and $X\in [N]^i$,
there exists a block in $\calS$ that covers $X$.
A covering design/Tur\'an system/SE system (with prescribed parameters)
is \emph{minimal} if it contains the least number of blocks.

By definition, an $(n,t)$-SE system is also an $(n,t+1,t)$-Tur\'an system.
Hence, we have
\begin{equation}\label{eqn4}
S(n,t)\geq T(n,t+1,t)\geq \frac{1}{n-t} {n\choose t+1} = \frac{1}{t+1} {n\choose t},
\end{equation}
where the second inequality \cite{stop:katona64} follows by noting that
each block in the Tur\'an system is contained in $(n-t)$ distinct $(t+1)$-subsets.

Let $\C$ be an $[n,k=n-d+1,d]$ linear MDS code. 
Recall that the stopping redundancy of $\C$ is related to SE and Tur\'an numbers
in the following way:
\[
T(n,d-1,d-2) \leq \rho(\C) \leq S(n,d-2),
\]
or, equivalently,
\[
T(n,n-k,n-k-1) \leq \rho(\C) \leq S(n,n-k-1).
\]

We now focus on upper bounds on $S(n,t)$.

\subsection{Probabilistic Bounds}

Let $N$ be an $n$-set.
Consider the following random experiment in which we build 
an $(n,t)$-SE system, $\calS\subseteq [N]^t$.
In the first step, for a prescribed real value $p\in [0,1]$, insert into $\calS$ 
each element of $[N]^{t}$ with probability $p$.
The expected size of $\calS$ at this point is
$
p {n\choose t},
$
but some $i$-subsets, $i=1,\ldots,t+1$, may not be covered.
The probability that a given $i$-subset is not covered equals
$
(1-p)^{\varphi(n,t,i)},
$
where
\[
\varphi(n,t,i)=i {n-i\choose t-i+1}=i{n-i\choose n-t-1}.
\]
So, as a second step, for each $X\in [N]^{i}$, $i=1,\ldots,t+1$, that
is not yet covered, insert into $\calS$ some element of $[N]^{t}$ that
covers $X$. 
The expected size of $\calS$ is then bounded from above by
\[
p{n\choose t} + \sum^{t+1}_{i=1} {n\choose i} (1-p)^{\varphi(n,t,i)}.
\]
This implies the following upper bound on $S(n,t)$.

\begin{theorem}\label{thm1}
For all $0\leq p\leq 1$, 
\begin{equation}\label{eqn3}
S(n,t)\leq
p{n\choose t} + \sum^{t+1}_{i=1} {n\choose i} (1-p)^{i {n-i\choose t-i+1}}.
\end{equation}
\end{theorem}

Alternatively, in the first step of the random experiment, we may 
instead make $l$ random drawings from $[N]^{t}$.
At the end of the first step,
the probability that a given $i$-subset is not covered equals
\[
\left(1-\frac{\varphi(n,t,i)}{{n\choose t}}\right)^l
\]
if the drawing is done with replacement, and equals
\[
\prod^{l-1}_{j=0}
\left(
1-\frac{\varphi(n,t,i)}{{n\choose t}-j}
\right)^+ ,
\]
where $(x)^+ =\max\{x,0\}$,
if the drawing is done without replacement.
The results are the following bounds.

\begin{theorem}\label{thm2}
For all $l\in\mathbb{N}$,
\begin{equation}\label{eqn1}
S(n,t)\leq 
l+\sum^{t+1}_{i=1} {n\choose i} 
\left(1-\frac{i {n-i\choose t-i+1}}{{n\choose t}}\right)^l.
\end{equation}
\end{theorem}

\begin{theorem}\label{thm3}
For all $l\in\mathbb{N}$, $l\leq {n\choose t}$,
\begin{equation}\label{eqn2}
S(n,t)\leq 
l+\sum^{t+1}_{i=1} {n\choose i}
\prod^{l-1}_{j=0} \left(1-\frac{i {n-i\choose t-i+1}}{{n\choose t}-j}\right)^+.
\end{equation}
\end{theorem}

Theorem~\ref{thm3} is clearly stronger than Theorem~\ref{thm2},
and is closely related to Theorem~\ref{thm1}.
In fact, one can show that
\[
\prod^{l-1}_{j=0} \left( 1- \frac{i {n-i\choose t-i+1}}{{n\choose t}-j}\right)^+
\leq
\left(1-\frac{l-1}{{n\choose t}}\right)^{i {n-i\choose t-i+1}}.
\]
So the upper bound \eqref{eqn2},
when minimized over $l$, is no greater than
\begin{equation*}
\min_{\substack{l\in\mathbb{Z}\\ 0\leq l\leq {n\choose t}}}
\left\{
l + \sum^{t+1}_{i=1} {n\choose i}
\biggl(1-\frac{l}{{n\choose t}}\biggr)^{\varphi(n,t,i)}
\right\}+1.
\end{equation*}
Note that we have strategically allowed $l$ to take the value ${n\choose t}$
in the above expression.
%
On the other hand, letting $l=p{n\choose t}$ in \eqref{eqn3},
we see the minimum value of the upper bound \eqref{eqn3} is
\begin{equation*}
\min_{\substack{l\in\mathbb{R}\\ 0\leq l\leq {n\choose t}}}
\left\{
l + \sum^{t+1}_{i=1} {n\choose i}
\biggl(1-\frac{l}{{n\choose t}}\biggr)^{\varphi(n,t,i)}
\right\}.
\end{equation*}
Now, suppose the minimum value of the above expression is $y$, achieved at $l=l^*$.
Then its value at $l=\ceil{l^*}$ is less than $y+1$.
Therefore, we conclude that the upper bound \eqref{eqn2} (when minimized over $l$)
is less than the upper bound \eqref{eqn3} (when minimized over $p$) plus two.
In practice, the difference between the two bounds is very small,
while the upper bound in \eqref{eqn3} is usually easier to compute.

The upper bound in \eqref{eqn3} can be written as
\begin{equation}
\label{eqn5}
\left(p+(1-p)^{t+1}\frac{n-t}{t+1}\right) {n\choose t}
+\sum^{t}_{i=1}(1-p)^{\varphi(n,t,i)} {n\choose i}.
\end{equation}
The first term in \eqref{eqn5} is minimized when $p$
takes the value
\[
p_{\min} =1-(n-t)^{-1/t},
\]
in which case \eqref{eqn5} becomes
\begin{equation}\label{eqn6}
\bigl(1-(n-t)^{-1/t}+\eta(n,t)\bigr)\cdot {n\choose t},
\end{equation}
where
\begin{align}
\eta(n,t)\nonumber
&= {n\choose t}^{-1}\cdot
                \sum^{t+1}_{i=1} (1-p_{\min})^{\varphi(n,t,i)} {n\choose i}\nonumber\\
& = \frac{n-t}{t+1}\cdot \sum^{t+1}_{i=1}
       \left((n-t)^{-\frac{i}{t}{n-i\choose n-t-1}}\right)
       \frac{{t+1\choose i}}{{n-i\choose n-t-1}}.\label{eqn7}
\end{align}
As $n\rightarrow\infty$, it can be shown that (see Appendix~\ref{app1})
if $t<n-\ln n$, then the term corresponding to $i=t+1$
prevails in the sum \eqref{eqn7}.
Therefore,
\[
\eta(n,t)=\bigl(1+o(1)\bigr)\cdot \frac{(n-t)^{-1/t}}{t+1}.
\]
Plugging the above into \eqref{eqn6}, we conclude that as $n\rightarrow\infty$,
the bound \eqref{eqn6} is%
\footnote{We write $f(n)\prec g(n)$, if $f(n)=o\bigl(g(n)\bigr)$,
and similarly, $f(n)\succ g(n)$, if $g(n)=o\bigl(f(n)\bigr)$, as $n\rightarrow\infty$.}
\[
\begin{cases}
\bigl(1+O(n^{-1/t})\bigr){n\choose t}, & \text{if $t\prec \ln n$}\\
\left(1- e^{-1/c}+o(1)\right){n\choose t}, & \text{if $t=\bigl(c+o(1)\bigr)\ln n$}\\
\Bigl(1+O\bigl(\frac{\ln (n-t)}{t}\bigr)\Bigr)\frac{\ln (n-t)}{t}{n\choose t}, & \text{if $\ln n \prec t <n-\ln n$}
\end{cases}
\]
where $c>0$ is any constant.
By the choice of $p_{\min}$, and the fact 
that for $p=p_{\min}$ the $(t+1)$-st term prevails in \eqref{eqn7}, 
no other values of $p$ give asymptotically tighter bounds than the above.

For the case when $t\geq n-\ln n$, a different asymptotic analysis
(see Appendix~\ref{app2}) shows that
the bound \eqref{eqn3}, when minimized over $p$, is
$O\bigl(\frac{\ln n}{n}{n\choose t}\bigr)$ for all $t$ such that
$2<n-t=o\bigl((n\ln\ln n)/\ln n\bigr)$, and is, particularly, 
$\Theta\bigl(\frac{\ln n}{n}{n\choose t}\bigr)$
for all $t=n-\Theta(1)$.



\subsection{Constructive Bounds}

Our first construction for SE systems is based on 
a Tur\'an system construction due to
Kim and Roush \cite{stop:kim83}. 

\begin{construction}\label{const1}
Let $N$ be an $n$-set, and $t<n-2$ be a positive integer.
For a prescribed positive integer $l$, 
partition $N$ into $l$ subsets, $N_i$, $i=0,\ldots,l-1$, as equally as possible. 
Thus, $N=\bigcup^{l-1}_{i=0} N_i$, such that 
$\lfloor n/l\rfloor \leq |N_i|\leq\lceil n/l\rceil$ for all $i$.
For all $X\subseteq N$, define the \emph{weight} of $X$ as
$w(X)=\sum^{l-1}_{i=0} i|X\cap N_i|$.
For each $j\in \{0,\ldots,l-1\}$, a subset of $[N]^t$ is chosen as
\[
B_j=Z\cup \tilde{B}_j,
\]
where
\[
Z=\{X\in [N]^t: \exists m, X\cap N_m=\emptyset, N_{m-1}\nsubseteq X\},
\]
and
\[
\tilde{B}_j=\{X\in [N]^t: w(X)\equiv j \mod l\}.
\]
Note that in the above definition of $Z$, the subscript $m-1$ is to be interpreted as 
$(m-1\mod l)$,
and we shall stick to this convention where applicable.
\end{construction}

\begin{proposition}\label{prop1}
For all $j$ and all $l\geq n/(n-t-2)$, $B_j$ as given in Construction~\ref{const1}
is an $(n,t)$-SE system.
\end{proposition}

\begin{proof}
We show that any $X\in [N]^i$, $i=1,\ldots,t+1$, is covered by a block in $B_j$.
If $X\cap N_m=\emptyset$ for some $m$, let $Y\in [N\setminus N_m]^{t+1}$ be selected
such that $X\subseteq Y$ and $|Y\cap N_{m-1}|$ is as small as possible.
Since $l\geq n/(n-t-2)$, we have $n- |N_m| \geq t+2$,
which ensures that $Y$ exists and 
that if $N_{m-1}\nsubseteq X$ then $N_{m-1}\nsubseteq Y$.
Now, choose $x\in X$ such that if $N_{m-1}\subseteq X$ then $x\in N_{m-1}$,
otherwise arbitrarily.
Note that $X$ is covered by $Y\setminus\{x\}$.
But we also have $(Y\setminus\{x\})\cap N_m=\emptyset$, and
$N_{m-1}\nsubseteq (Y\setminus\{x\})$.
Therefore, $Y\setminus\{x\}\in Z$. 

On the other hand, if $X\cap N_m\neq \emptyset$ for all $m$, select one element in 
each such intersection, say $x_m\in X\cap N_m$.
Now, choose $Y\in [N]^{t+1}$ such that $X\subseteq Y$, 
and consider $Y\setminus\{x_m\}$, $m=0,\ldots,l-1$.
All these sets cover $X$, and since $w(Y\setminus\{x_m\})=w(Y)-m$, 
they have distinct weights that span $l$ consecutive integers,
one of which must be congruent to $j$ modulo $l$.
Hence, for all $j$, there exists $m$ such that $Y\setminus\{x_m\}\in \tilde{B}_j$.
\end{proof}

\begin{theorem}\label{thm4}
For all integers $l\geq n/(n-t-2)$, 
\[
S(n,t)\leq
\frac{1}{l} {n\choose t} +
l\left[{n-\lfloor\frac{n}{l}\rfloor\choose t} - 
{n-\lfloor\frac{n}{l}\rfloor-\lceil\frac{n}{l}\rceil \choose 
t-\lceil\frac{n}{l}\rceil}\right].
\]
\end{theorem}

\begin{proof}
From Proposition~\ref{prop1}, for all $l\geq n/(n-t-2)$, 
\[
S(n,t)\leq \min_{j} |B_j|.
\]
Note that
\[
Z=\bigcup^{l-1}_{m=0} \bigl(Z_{1,m}\setminus Z_{2,m}\bigr),
\]
where
\[
Z_{1,m}=\{X\in [N]^t: X\cap N_m=\emptyset\},
\]
\[
Z_{2,m}=\{X\in [N]^t: X\cap N_m=\emptyset, N_{m-1}\subseteq X\}.
\]
Also, 
\[
\bigcup^{l-1}_{j=0} \tilde{B}_j = [N]^t.
\]
Thus, we have
\begin{align*}
&\min_{j} |B_j|\\
& = |Z| + \min_j |\tilde{B}_j|\\
& \leq \sum^{l-1}_{m=0} \left[{n-|N_m|\choose t} -
{n-|N_m|-|N_{m-1}|\choose t-|N_{m-1}|}\right]
+ \min_j |\tilde{B}_j|\\
& \leq l \left[{n-\lfloor\frac{n}{l}\rfloor\choose t} -
{n-\lfloor\frac{n}{l}\rfloor-\lceil\frac{n}{l}\rceil \choose
t-\lceil\frac{n}{l}\rceil}\right]
+ \frac{1}{l} {n\choose t}.
\end{align*}
\end{proof}

An alternative (slightly looser) form of the upper bound is given 
in the following theorem.

\begin{theorem}\label{thm5}
For all integers $l\geq n/(n-t-2)$,
\begin{equation*}
S(n,t)\leq
\frac{1}{l} {n\choose t} +
l\left\lceil\frac{n}{l}\right\rceil
{n-\lfloor\frac{n}{l}\rfloor-1 \choose t}.
\end{equation*}
\end{theorem}

\begin{proof}
Note that 
\[
Z=\bigcup^{l-1}_{m=0} \bigcup_{\alpha\in N_{m-1}} 
\Bigl[N\setminus \bigl(N_m\cup\{\alpha\}\bigr)\Bigr]^t.
\]
The rest of the proof is similar to that of Theorem~\ref{thm4}.
\end{proof}

\begin{corollary}\label{cor4}
For fixed $k$, as $n\rightarrow\infty$,
\[
S(n,n-k-1)\leq \left(\frac{2}{k+1}+O(n^{-1})\right) {n\choose k}.
\]
\end{corollary}

\begin{proof}
Theorem~\ref{thm5} applies provided that $l\geq n/(k-1)$.
If $k\geq 4$, let $l=\lfloor n/2\rfloor$.
We have
\begin{align*}
&S(n,n-k-1)\\
&\leq \frac{2}{n}\left(1+O\left(\frac{1}{n}\right)\right) {n\choose n-k-1} + 
  \frac{3n}{2} {n-3\choose n-k-1}\\
&= \frac{2}{k+1} {n\choose k} + O(n^{k-1}).
\end{align*}

For $k=3$, let $l=\ceil{n/2}$.
If $n$ is even, the above derivation is still valid.
If $n$ is odd, note that there is one bin that contains a single element, and
the rest $(n-1)/2$ bins all contain two elements.
From the proof of Theorem~\ref{thm5}, we have
\begin{align*}
|Z|
&\leq \sum^{\ceil{n/2}-1}_{m=0} |N_{m-1}| \cdot {n-|N_m|-1\choose n-4}\\
&= (n-2) \cdot {n-3\choose n-4} + 2\cdot {n-2\choose n-4}\\
&= O\bigl(n^2\bigr)
\end{align*}
Hence,
\begin{align*}
S(n,n-4)
&\leq \frac{2}{n+1} {n\choose n-4} + O\bigl(n^2\bigr)\\
&= \frac{1}{2} {n\choose 3} + O\bigl(n^2\bigr)
\end{align*}

For $k<3$, the result has already been shown in \cite{stop:han07a}.
\end{proof}

Since
\[
T(n,n-k,n-k-1)\geq \frac{1}{k+1}{n\choose k},
\]
Corollary~\ref{cor4} also implies that for any fixed $k$,
\[
S(n,n-k-1)\leq \bigl(2+O(n^{-1})\bigr) T(n,n-k,n-k-1),
\]
confirming a conjecture made in \cite{stop:han07a}.
Note that the asymptotic factor of $2$ in the above inequality is sharp
for $k=1$, in which case $S(n,n-2)=n-1$, while $T(n,n-1,n-2)=\lceil n/2 \rceil$.
For $k>1$, stronger results can be obtained using recurrent inequalities, 
as shall be discussed in the next section.

Construction~\ref{const1} is also a construction for Tur\'an systems. 
Indeed, it can be viewed as an improved version
(i.e. one with fewer blocks) of the Tur\'an system
construction in \cite{stop:kim83}.

\begin{proposition}
For all $j$ and all $l$, $B_j$ as given in Construction~\ref{const1}
is a Tur\'an $(n,t+1,t)$-system.
\end{proposition}

\begin{proof}
The proof is similar to that of Proposition~\ref{prop1}.
\end{proof}

Our second construction for SE systems is
inspired by a construction for Tur\'an systems 
due to Frankl and R\"odl \cite{stop:frankl85}.

\begin{construction}\label{const2}
Let $N$ be an $n$-set, and $t<n$ be a positive integer.
Let $N_i$, $i=0,\ldots,l-1$, and $w(X)$, $X\subseteq N$, 
be defined as in Construction~\ref{const1}.
We will call $N_i$ a \emph{bin}.
For each $j\in \{0,\ldots,l-1\}$, let
\[
\tilde{B}_j=\bigl\{X\in [N]^t: w_j(X) \leq \max\{e(X),f(X)\}\bigr\},
\]
where
\[
w_j(X)=(w(X)+j)\mod l,
\]
and
\[
e(X)=|\{i:X\cap N_i=\emptyset\}|,
\]
\[
f(X)=|\{i:N_i\subseteq X\}|
\]
are the number of ``empty'' and ``full'' bins for $X$, respectively.
The constructed collection of $t$-subsets of $N$ is
\[
B_j=F\cup \tilde{B}_j,
\]
where $F$ is constructed as follows.

Fix an arbitrary total order on $N$.
Let $I \subseteq \{0,\ldots,l-1\}$ be an index set that
satisfies $\sum_{m\in I} |N_m| > t$, and is \emph{minimal} in the sense that all proper
subsets of $I$ violate this condition. For each such $I$ and
$i,j\in I$, $i\neq j$, let $F$ include the $t$-set
that contains all elements from bins $N_m, m \in I \setminus \{i,j\}$, 
the smallest $|N_i|-1$ elements of $N_i$,
and the smallest $\bigl(t+1-\sum_{m\in I, m\neq j} |N_m|\bigr)$ elements of $N_j$.
\end{construction}

\begin{proposition}
For all $j$ and all $l$, $B_j$ as given in Construction~\ref{const2}
is an $(n,t)$-SE system.
\end{proposition}

\begin{proof}
We show that any $X\in [N]^i$, $i=1,\ldots,t+1$, is covered by a block in $B_j$.
If $i=t+1$, note that all $t$-subsets of $X$ can be written as $X\setminus\{x\}$,
for some $x\in X$. 
Since $w(X\setminus\{x\})=w(X)-w(\{x\})$ for all $x\in X$,
by choosing $x$ from different bins 
that $X$ intersects, we can make $w(X\setminus\{x\})$ take on $l-e(X)$ different values.
Since no two of these values differ by more than $l-1$, this also means that we can
realize $l-e(X)$ different values for $w_j(X\setminus\{x\})$.
Since only $l-e(X)-1$ numbers in $\{0,\ldots,l-1\}$ 
are greater than $e(X)$, there exists $x\in X$ such that 
$w_j(X\setminus\{x\})\leq e(X)\leq e(X\setminus\{x\})$, hence
$X\setminus\{x\}\in \tilde{B}_j$, and it covers $X$.

If $i\leq t$, consider two cases. 
First, let us assume that there exists $m$, 
such that $X\cap N_m\neq\emptyset$ and $N_m\nsubseteq X$. 
In this case, remove from $X$ an arbitrary element in $X\cap N_m$, 
add in $t-i$ other elements from $N$ using as few elements from $N_m$ as possible, 
and call the resulting $(t-1)$-set $\tilde{X}$.
That is, $\tilde{X}=(X\setminus\{x\})\cup Y$, for some $x\in X\cap N_m$ and some
$Y\in [N\setminus X]^{t-i}$ that has a minimal number of elements from $N_m$.
Note that the choice of $\tilde{X}$ ensures that $f(\tilde{X}\cup\{x\})=f(\tilde{X})$.
Since $w(\tilde{X}\cup\{z\})=w(\tilde{X})+w(\{z\})$ for all $z\notin\tilde{X}$,
by choosing $z\notin\tilde{X}$, $z\neq x$, from different bins where possible,
we can make $w(\tilde{X}\cup\{z\})$ take on 
$l-f(\tilde{X}\cup\{x\})=l-f(\tilde{X})$ different values.
This also means that we can realize $l-f(\tilde{X})$ different values 
for $w_j(\tilde{X}\cup\{z\})$.
Since only $l-f(\tilde{X})-1$ numbers in $\{0,\ldots,l-1\}$
are greater than $f(\tilde{X})$, there exists $z$ such that
$w_j(\tilde{X}\cup\{z\})\leq f(\tilde{X})\leq f(\tilde{X}\cup\{z\})$, hence
$\tilde{X}\cup\{z\}\in \tilde{B}_j$, and it covers $X$.

Next, if no $m$ exists such that $X\cap N_m\neq\emptyset$ and $N_m\nsubseteq X$,
this means that for all $m$ such that $X\cap N_m\neq\emptyset$, we have
$N_m\subseteq X$. 
Figuratively, it means that $X$ consists of a number of full bins.
Let $N_i$ be any bin that $X$ intersects. 
Let $x\in N_i$ be its largest element. 
Take $X\setminus\{x\}$, and add to it elements from bins that $X$ does not intersect, 
one bin after another, from smallest to the largest within each bin, until 
$X\setminus\{x\}$ is augmented to contain $t$ elements.
By construction, the $t$-subset thus obtained is contained in $F$.
\end{proof}

\begin{theorem}\label{thm9}
For all positive integers $l$, 
\[
S(n,t)\leq
\frac{1}{l} {n\choose t} +
{n-\lfloor\frac{n}{l}\rfloor\choose t} +
{n-\lfloor\frac{n}{l}\rfloor\choose t-\lfloor\frac{n}{l}\rfloor}
+ g(n,t,l),
\]
where
\[
g(n,t,l)=\sum_{\frac{t+1}{\lceil n/l\rceil}\leq i\leq \left\lceil\frac{t+1}{\lfloor n/l\rfloor}\right\rceil}
  {l\choose i} i(i-1).
\]
\end{theorem}

\begin{proof}
Note that 
\begin{align*}
\sum_{X\in [N]^t} f(X) & = \sum_{X\in [N]^t} \sum_i 1_{\{N_i\subseteq X\}}\\
& = \sum_i \Bigl(\sum_{X\in [N]^t} 1_{\{N_i\subseteq X\}}\Bigr)\\
& \leq l {n-\lfloor\frac{n}{l}\rfloor\choose t-\lfloor\frac{n}{l}\rfloor}.
\end{align*}
Similarly,
\[
\sum_{X\in [N]^t} e(X) 
\leq 
l {n-\lfloor\frac{n}{l}\rfloor\choose t}.
\]
Each $X\in [N]^t$ is contained in precisely 
$1+\max\{e(X),f(X)\}$ $\tilde{B}_j$'s.
Therefore,
\begin{align*}
\sum_j |\tilde{B}_j| & = \sum_{X\in [N]^t} \bigl(1+\max\{e(X),f(X)\}\bigr)\\
& \leq {n\choose t} + \sum_{X\in [N]^t} \bigl(e(X)+f(X)\bigr)\\
& \leq {n\choose t} + l {n-\lfloor\frac{n}{l}\rfloor\choose t}
  + l {n-\lfloor\frac{n}{l}\rfloor\choose t-\lfloor\frac{n}{l}\rfloor}.
\end{align*}
Hence,
\[
\min_j |\tilde{B}_j|
\leq
\frac{1}{l} {n\choose t} +
{n-\lfloor\frac{n}{l}\rfloor\choose t} +
{n-\lfloor\frac{n}{l}\rfloor\choose t-\lfloor\frac{n}{l}\rfloor}.
\]
Finally, note that for each valid $I$, $F$ contains $|I|(|I|-1)$ $t$-subsets, 
and a valid $I$ must satisfy $|I|\lceil n/l\rceil \geq t+1$ and 
$(|I|-1)\lfloor n/l\rfloor < t+1$.
Therefore, $|F|\leq g(n,t,l)$.
\end{proof}

\subsection{Recurrent Bounds}

We observe that an $(n,t)$-SE system can be constructed from
an $(n-1,t-1)$-SE system and an $(n-1,t+1,t)$-Tur\'an system,
as shown in the following lemma.

\begin{lemma}\label{lem1}
For all $0<t<n-1$,
\[
S(n,t)\leq S(n-1,t-1)+T(n-1,t+1,t),
\]
or equivalently, for all $0<k<n-1$,
\[
S(n,n-k-1)\leq S(n-1,n-k-2)+C(n-1,k,k-1).
\]
\end{lemma}

\begin{proof}
Let $N$ be an $n$-set and $a\in N$ be an arbitrary element.
Let $\calS\subseteq [N\setminus\{a\}]^{t-1}$ be a minimal $(n-1,t-1)$-SE system,
and $\T\subseteq [N\setminus\{a\}]^{t}$ be a minimal $(n-1,t+1,t)$-Tur\'an system.
Define $\calS'=\{s\cup\{a\}:s\in\calS\}$.
Then $\calS'\cup\T$ is an $(n,t)$-SE system.
Indeed, for all $X\in [N]^i$, $i=1,\ldots,t+1$,
if $1\leq |X\setminus\{a\}|\leq t$, then
there exists $s\in\calS$ such that $|(X\setminus\{a\})\setminus s|=1$, 
which implies that $|X\setminus (s\cup\{a\})|=1$, 
i.e. $X$ is covered by a block in $\calS'$. 
The only cases left are when $X=\{a\}$, 
and when $X\in [N\setminus\{a\}]^{t+1}$. 
In either case, $X$ is covered by a block in $\T$.
\end{proof}

\begin{theorem}\label{thm6}
For all $0<t<n-1$,
\begin{equation}\label{eqn10}
S(n,t)\leq \sum^{t}_{i=0} T(n-t+i-1,i+1,i),
\end{equation}
or equivalently, for all $0<k<n-1$,
\begin{equation}\label{eqn9}
S(n,n-k-1)\leq \sum^{n-1}_{i=k} C(i,k,k-1).
\end{equation}
\end{theorem}

\begin{proof}
Recursively apply Lemma~\ref{lem1}.
\end{proof}

Interesting results follow.
When $k=1$, \eqref{eqn9} implies that $S(n,n-2)\leq n-1$, which is sharp.
When $k=2$, since $C(i,2,1)=\lceil i/2\rceil$, \eqref{eqn9} gives
\[
S(n,n-3)\leq \left\lceil\frac{n}{2}\right\rceil \left\lfloor\frac{n}{2}\right\rfloor - 1,
\]
which is asymptotically tighter than Corollary~\ref{cor4}.

Generally, since exact values of most Tur\'an / covering numbers are not known, 
the upper bounds in Theorem~\ref{thm6} often cannot be directly evaluated. 
To get a computable upper bound, one can simply replace each Tur\'an / covering
number in the sum by an explicit upper bound. 
We show several ways to do this.
The first one is based on a result by Erd\H{o}s and Spencer \cite{stop:erdos74}.

\begin{theorem}\label{thm7}
For all $0<k<n-1$,
\[
S(n,n-k-1)\leq \frac{1+\ln k}{k} \left({n\choose k}-1\right).
\]
\end{theorem}

\begin{proof}
In \cite{stop:erdos74}, it was shown that for all $n\geq s\geq t$,

\begin{equation}\label{eqn11}
C(n,s,t)\leq \left(1+\ln {s\choose t}\right) \frac{{n\choose t}}{{s\choose t}}.
\end{equation}

Plugging \eqref{eqn11} into \eqref{eqn9}, we obtain the claimed result
after some algebraic manipulations.
\end{proof}

The second one is based on an upper bound on Tur\'an numbers 
due to Sidorenko \cite[Construction~4]{stop:sidorenko97}.

\begin{theorem}\label{thm10}
For all $0<t<n-1$ and positive integers $l_0,l_1,\ldots,l_t$,
\[
S(n,t)\leq \sum^{t}_{i=0} f_{n,t}(i,l_i),
\]
where 
\begin{multline*}
\hspace{-5pt}f_{n,t}(i,l_i)=\\
\left[\frac{1}{2l_i}+\frac{1}{2}\hspace{-2pt}
\left(\hspace{-2pt}3+\frac{i}{l_i-1-\frac{l_i(i-1)}{m+i}}\hspace{-2pt}\right)
\hspace{-2pt}\left(1-\frac{1}{l_i}\right)^{\hspace{-1pt}i}\right]
\hspace{-1.5pt}\cdot\hspace{-1.5pt} {m+i\choose i},\hspace{-5pt}
\end{multline*}
and $m=n-t-1$.
\end{theorem}

\begin{proof}
Omitted.
\end{proof}

A third way to obtain an explicit upper bound from Theorem~\ref{thm6}
is based on a construction of an $(n,k,k-1)$ covering design due to
Kuzjurin \cite{stop:kuzjurin77}, although we count blocks in a slightly different manner.

\begin{lemma}
For all positive integers $n\geq k$,
\[
C(n,k,k-1)\leq \frac{1}{k} {n\choose k-1} + \frac{k-1}{k} {n-1\choose k-2}.
\]
\end{lemma}

\begin{proof}
Let $N$ be an $n$-set. 
WLOG, let $N=\{1,2,\ldots,n\}$.
Let $Q_i=\{X\in[N]^k: \sum_{x\in X} x \equiv i \mod n\}$,
and $C_i=\{X\in[N]^{k-1}: \nexists Y\in Q_i, \mbox{ s.t. } X\subset Y\}$,
$i=0,1,\ldots,n-1$.
For each $X\in C_i$, we can add one block $Y\in[N]^k$ to $Q_i$,
such that $X\subset Y$.
Hence, by adding no more than $|C_i|$ blocks to $Q_i$, we construct
an $(n,k,k-1)$ covering design. 
Therefore, for all $i$,
\[
C(n,k,k-1)\leq |Q_i| + |C_i|.
\]
Now note that for all $Y, Z\in Q_i$, $Y\neq Z$, we have $|Y\cap Z|\leq k-2$.
Therefore, for all $X\in [N]^{k-1}$ there is at most one block $Y$ in $Q_i$
such that $X\subset Y$;
on the other hand, for every $Y\in [N]^k$ there are $k$ elements $X\in [N]^{k-1}$
such that $X\subset Y$.
Hence,
\[
|C_i|={n\choose k-1} - k |Q_i|.
\]
We have
\begin{align*}
\sum^{n-1}_{i=0} \bigl( |Q_i|+|C_i| \bigr)
&= \sum^{n-1}_{i=0} \left({n\choose k-1} - (k-1)|Q_i|\right)\\
&=n {n\choose k-1} - (k-1) {n\choose k}.
\end{align*}
Therefore, there exists $i$, such that
\begin{align*}
|Q_i|+|C_i|
&\leq {n\choose k-1} - \frac{k-1}{n} {n\choose k}\\
&= \frac{1}{k} {n\choose k-1} + \frac{k-1}{k} {n-1\choose k-2}.
\end{align*}
\end{proof}

Plugging the bound in the preceding lemma into \eqref{eqn9}, we obtain the following theorem.

\begin{theorem}\label{thm11}
For all $0<k<n-1$,
\[
S(n,n-k-1)\leq \frac{1}{k} {n\choose k} + \frac{k-1}{k} {n-1\choose k-1} -1.
\]
\end{theorem}

\begin{proof}
Omitted.
\end{proof}

\begin{corollary}\label{cor1}
For all $0<k=o(\sqrt{n})$, as $n\rightarrow\infty$, 
\[
S(n,n-k-1)\leq \bigl(1+o(1)\bigr) \frac{1}{k} {n\choose k}.
\]
In particular, if in addition $k\rightarrow\infty$, then
\[
S(n,n-k-1)\sim T(n,n-k,n-k-1).
\]
\end{corollary}

\begin{proof}
Omitted.
\end{proof}

Interestingly, a similar asymptotic result can be shown for $S(n,t)$
when $t$ is relatively small. 
First, we note the following theorem relating SE and Tur\'an numbers.

\begin{theorem}\label{thm8}
For all $0<t<n-1$,
\[
S(n,t)\leq \left(1-\frac{t}{n}\right) T(n,t+1,t) + {n-1\choose t-1}.
\]
\end{theorem}

\begin{proof}
We will use the fact \cite{stop:schoenheim64} that
\[
T(n,s,t)\geq \frac{n}{n-t} T(n-1,s,t).
\]
From Theorem~\ref{thm6}, we have
\begin{align*}
S(n,t)&\leq T(n-1,t+1,t) + \sum^{t-1}_{i=0} T(n-t+i-1,i+1,i)\\
&\leq \left(1-\frac{t}{n}\right)T(n,t+1,t) + \sum^{t-1}_{i=0} {n-t+i-1\choose i}\\
&= \left(1-\frac{t}{n}\right)T(n,t+1,t) + \sum^{n-2}_{i=n-t-1} {i\choose n-t-1}\\
&= \left(1-\frac{t}{n}\right)T(n,t+1,t) + {n-1\choose t-1}.
\end{align*}
\end{proof}

In \cite[Theorem~21]{stop:han07a}, it was shown that $S(n,t)$ is asymptotic to
$T(n,t+1,t)$ for any fixed $t$ as $n\rightarrow\infty$.
Theorem~\ref{thm8} enables us to extend this result to all $t=o(\sqrt{n})$.

\begin{corollary}\label{cor2}
For all $t=o(\sqrt{n})$, as $n\rightarrow\infty$, 
\[
S(n,t)\sim T(n,t+1,t).
\]
\end{corollary}

\begin{proof}
Note that
\[
\frac{{n-1\choose t-1}}{T(n,t+1,t)}
\leq \frac{{n-1\choose t-1}}{\frac{1}{t+1}{n\choose t}}
= \frac{t^2+t}{n}.
\]
The result then follows immediately from Theorem~\ref{thm8}.
\end{proof}

\subsection{Comparison of Upper Bounds}

\begin{figure}
\centering
\includegraphics[width=\figscale\columnwidth]{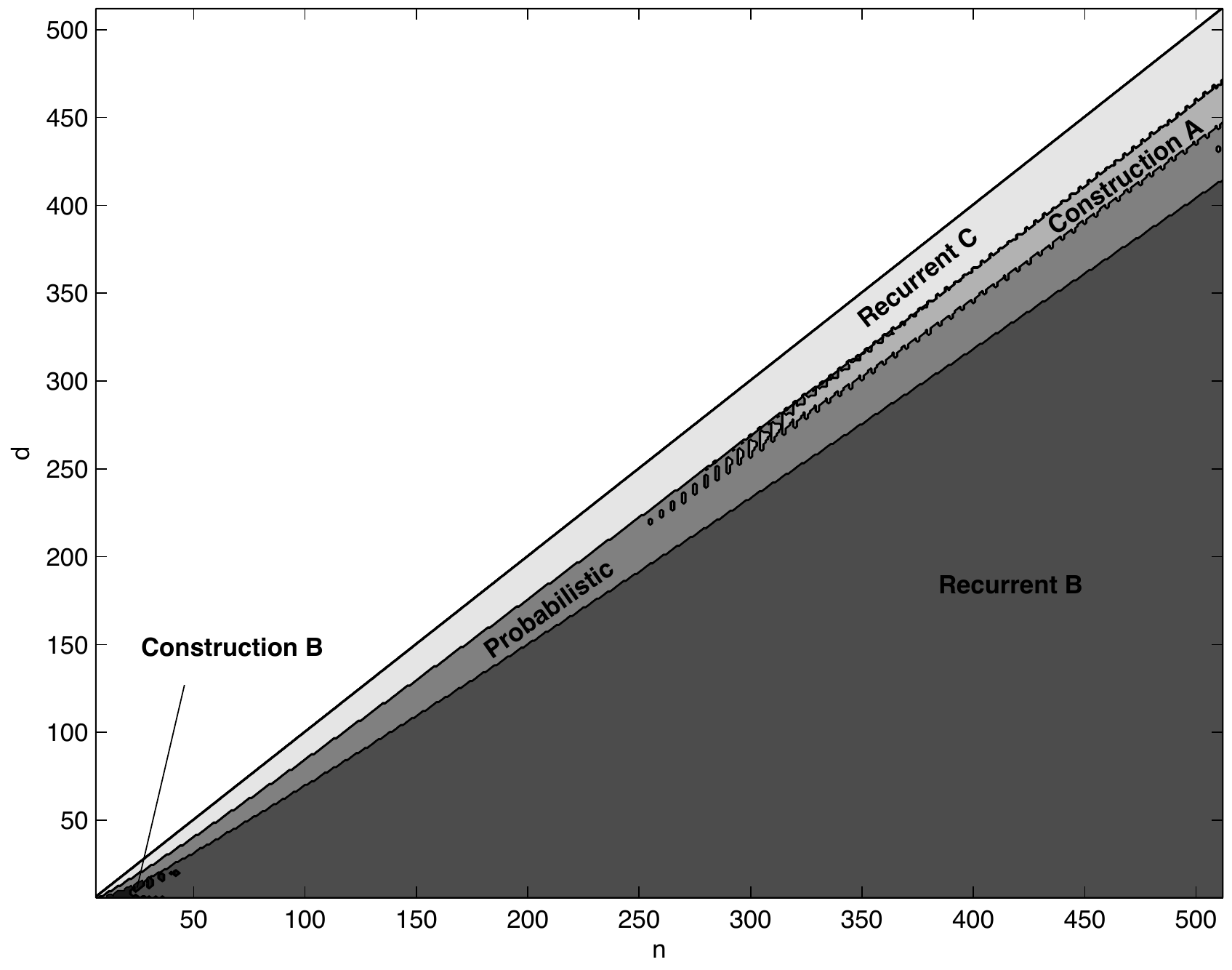}
\caption{Best upper bounds on $S(n,d-2)$ (hence on $\rho(\C)$), for $5< d\leq n\leq 512$.}
\label{fig1}
\end{figure}

We numerically computed several of the upper bounds on $S(n,d-2)$, and hence on $\rho(\C)$, for all $5< d\leq n\leq 512$. 
For each $(n,d)$ pair, the tightest bound is identified,
and the results are shown in Fig.~\ref{fig1}.
In the figure,
``Construction~A'' refers to Theorem~\ref{thm4},
``Construction~B'' refers to Theorem~\ref{thm9},
``Probabilistic'' refers to Theorem~\ref{thm1},
``Recurrent B'' refers to Theorem~\ref{thm10},
and ``Recurrent C'' refers to Theorem~\ref{thm11}.
Other bounds included in the comparison (but not appearing in the figures) are
``Schwartz-Vardy'' \eqref{eqn0},
``Construction~1'' (\cite[Theorem~27]{stop:han07a}),
``Construction~3'' (\cite[Theorem~39]{stop:han07a}),
and ``Recurrent~A'' (Theorem~\ref{thm7}).
Note that $d\leq 5$ is not considered, 
since in this case $S(n,d-2)$ (and $\rho(\C)$)
is known to be at most $T(n,d-1,d-2)+1$, for which either precise formulas
are known, or tighter special upper bounds exist \cite{stop:han07a}.

We observe that the most ``successful'' bounds are Recurrent B, Recurrent C,
Probabilistic, and Construction A.
A minor exception is Construction B, which excels
occasionally for certain small values of $(n,d)$.
As $n$ gets larger, a trend can be seen.
Roughly speaking, for code rate $(n-d+1)/n\geq 1/5$, Recurrent B is
the best bound.
For code rate less than $1/5$, the best bounds are Probabilistic,
Construction A, and Recurrent C, in that respective order, 
as code rate gets progressively smaller.

A few samples of the upper bounds are given in Table~\ref{tab1}.
The tightest are highlighted in boldface.
For comparison, a lower bound on $T(n,d-1,d-2)$ (hence on $S(n,d-2)$
and $\rho(\C)$ as well) has been included, based on \eqref{eqn13}
(see Section~\ref{sec2}).
Compared to upper bounds previously known, significant improvements
can clearly be seen.
As a side note, we caution that while Recurrent C is an excellent bound
when $d$ is very close to $n$,
it gets loose quickly as $d$ gets smaller, and should be avoided
if the code rate is greater than $1/2$.

\begin{table}
\renewcommand{\arraystretch}{1.1}
\caption{Upper bounds on $S(n,d-2)$ and $\rho(\C)$}
\label{tab1}
\begin{center}
\begin{tabular}{l|rrr}\hline\hline
$(n,d)$ & $(31,7)$ & $(31,23)$ & $(31,27)$ \\\hline
Probabilistic  &96,112 &\textbf{6,412,596} &77,298 \\
Construction A  &93,691 &7,693,683 &86,148 \\
Construction B  &76,986 &12,151,903 &299,697 \\
Recurrent A  &124,250 &7,161,809 &88,673 \\
Recurrent B  &\textbf{71,891} &9,665,343 &520,847 \\
Recurrent C  &599,474 &7,442,607 &\textbf{55,905} \\
Construction 1 \cite{stop:han07a}  &93,691 &7,786,707 &106,388 \\
Construction 3 \cite{stop:han07a} &76,986 &16,275,110 &269,970 \\
Schwartz-Vardy \eqref{eqn0}  &142,506 &31,475,730 &617,526 \\
Lower Bound \eqref{eqn13} &33,981 &2,103,660 &29,450 \\\hline\hline
\end{tabular}
\end{center}
\end{table}

\section{Lower Bounds on $S(n,t)$}\label{sec2}

A few other lower bounds on Tur\'an / covering numbers are known, besides
the simple lower bound in \eqref{eqn4}. 
For example, Sch\"onheim \cite{stop:schoenheim64} showed that
\begin{equation}\label{eqn13}
T(n,t+1,t)\geq \left\lceil \frac{n}{n-t}\left\lceil\frac{n-1}{n-t-1}\left\lceil\ldots\left\lceil\frac{t+2}{2}
\right\rceil\ldots\right\rceil\right\rceil\right\rceil.
\end{equation}
Another useful bound is due to De Caen \cite{stop:decaen83}:
\[
T(n,t+1,t)\geq \frac{1}{t}\cdot \frac{n-t}{n-t+1} {n\choose t}.
\]
Note that a lower bound on $T(n,d-1,d-2)$ is in turn a lower bound on $\rho(\C)$
and on $S(n,d-2)$.

In this section, we study a lower bound result on $S(n,t)$ (which is not 
a lower bound on $T(n,t+1,t)$ in general).
Note that a lower bound (just) on $S(n,d-2)$ 
is not necessarily a lower bound on $\rho(\C)$. 

\begin{theorem}
For all $0<t<n-1$,
\begin{align*}
&\left(1+ \frac{n-t}{n(n-t-1/2)}\right) S(n,t)\\
&\qquad\qquad\qquad\geq T(n-1,t+1,t) + \frac{1}{n-t-1/2}{n-1\choose t}.
\end{align*}
\end{theorem}

\begin{proof}
Let $N$ be an $n$-set, and $\calS\subseteq [N]^t$ be a minimal $(n,t)$-SE system.
For each $j\in N$, $\calS$ can be partitioned into blocks that contain $j$ and those
that do not, namely,
\[
\calS=\A_j \cup \B_j,
\]
where $\A_j=\{X\in\calS: j\in X\}$, and 
$\B_j=\{X\in\calS: j\notin X\}$. 
Note that for all $j$, $\B_j$ is an $(n-1,t+1,t)$-Tur\'an system.
Further, if we let $\A'_j=\{A\setminus\{j\}:A\in\A_j\}$,
then each $t$-set in $[N\setminus\{j\}]^t\setminus\B_j$ contains at least
one element of $\A'_j$. 
To see this, suppose $X\in[N\setminus\{j\}]^t\setminus\B_j$.
Then $X\cup\{j\}$ is a $(t+1)$-set and so
there exists $Y\in\calS$ such that $Y\subset \bigl(X\cup\{j\}\bigr)$. 
Since $X\notin \calS$, we have $j\in Y$ and hence 
$\A'_j\ni(Y\setminus\{j\})\subset X$.

On the other hand, since each element of $\A'_j$ is contained in
\[
(n-1)-(t-1)=n-t
\]
$t$-subsets of $N\setminus\{j\}$, it is contained in at most $(n-t)$ distinct
$t$-sets in $[N\setminus\{j\}]^t\setminus\B_j$.
Therefore,
\[
|\A_j|=|\A'_j|\geq \frac{1}{n-t} \left({n-1\choose t}-|\B_j|\right).
\]
This lower estimate can be improved by a more careful argument as follows.
Let
\begin{align*}
\A'_{j1}&=\{A\in\A'_j: \exists B\in\B_j, A\subset B\},\\
\A'_{j2}&=\{A\in\A'_j\setminus\A'_{j1}: \exists A'\in\A'_{j1}, |A\setminus A'|=1\},\\
\A'_{j3}&=\{A\in\A'_j\setminus(\A'_{j1}\cup\A'_{j2}): \exists A'\in\A'_{j2}, |A\setminus A'|=1\},\\
&\ldots\\
\A'_{ji}&=\biggl\{A\in\A'_j\setminus\bigcup^{i-1}_{l=1}\A'_{jl}: \exists A'\in\A'_{j(i-1)}, |A\setminus A'|=1\biggr\},\\
&\ldots
\end{align*}
Note that $\A'_{ji}\cap\A'_{jl}=\emptyset$ for all $i\neq l$.
Since $\A'_j$ is finite, there exists $i$ such that
$\A'_{jl}=\emptyset$ for all $l>i$.
Regardless, define
\[
\tilde{\A}'_j=\bigcup^{\infty}_{l=1} \A'_{jl}.
\]
We claim that on average, an element in $\tilde{\A}'_j$ is contained in at most $(n-t-1)$
~$t$-sets in $[N\setminus\{j\}]^t\setminus\B_j$.
To see this, consider a process in which we enumerate elements of $\A'_{j}$,
and for each element, ``mark'' the $t$-sets in $[N\setminus\{j\}]^t\setminus\B_j$
that contain it.
We start with elements in $\A'_{j1}$ and proceed to $\A'_{j2},\A'_{j3}$, and so on.
Each element in $\A'_{j1}$ is contained in $(n-t)$~~$t$-sets in 
$[N\setminus\{j\}]^t$, at least one of which lies in $\B_j$.
Therefore, for each element in $\A'_{j1}$, at most $(n-t-1)$~~$t$-sets are marked.
Now, for any $X\in\A'_{j2}$,
by definition, there exists $Y\in\A'_{j1}$,
such that $|X\setminus Y|=1$. 
Hence, among the $(n-t)$~~$t$-sets that contain $X$, at least one of them,
namely, $X\cup Y$, is already marked.
Therefore, processing any $X\in\A'_{j2}$ marks at most $(n-t-1)$ additional
$t$-sets in $[N\setminus\{j\}]^t\setminus\B_j$.
A similar argument shows that among the $(n-t)$~~$t$-sets that contain
an element of $\A'_{ji}$, at least one of them is already marked
after elements of $\A'_{j(i-1)}$ have been processed.

For $\A'_j\setminus\tilde{\A}'_j$, we show that on average, 
each element marks at most $(n-t-1/2)$~~$t$-sets in 
$[N\setminus\{j\}]^t\setminus\B_j$.
Let $X\in\A'_j\setminus\tilde{\A}'_j$.
As $X\cup\{j\}\in [N]^t$, there exists $Y\in\calS$, 
such that $|(X\cup\{j\})\setminus Y|=1$.
Since $X\notin\A'_{j1}$, we have $j\in Y$ and hence $Y\setminus\{j\}\in\A'_j$.
In fact, $Y\setminus\{j\}\in\A'_j\setminus\tilde{\A}'_j$,
since otherwise it would imply that $X\in\tilde{\A}'_j$.
Now, let $Z=X\cup Y\setminus\{j\}$, and denote by $l$ the number of elements
of $\A'_j\setminus\tilde{\A}'_j$ that are contained in $Z$.
Note that $l\geq 2$, since $Z$ contains both 
$X$ and $Y\setminus\{j\}$.
Therefore, of the $l$ elements that are contained in $Z$, each on average marks
\[
n-t-(l-1)/l\leq n-t-1/2
\]
$t$-sets in $[N\setminus\{j\}]^t\setminus\B_j$.
For other elements in $\A'_j\setminus\tilde{\A}'_j$,
the above argument can be repeated until all elements have been considered.

Based on the preceding discussions, we conclude that 
on average, each block in $\A'_j$ is contained 
in no more than $(n-t-1/2)$~~$t$-sets in $[N\setminus\{j\}]^t\setminus\B_j$.
Hence,
\begin{align}
|\calS|&=|\A'_j|+|\B_j|\nonumber\\
&\geq \frac{1}{n-t-1/2}\left({n-1\choose t}-|\B_j|\right)+T(n-1,t+1,t).\label{eqn12}
\end{align}
Since each block of $\calS$ appears in $(n-t)$~~$\B_j$'s,
we have
\[
\sum_{j\in N} |\B_j| = (n-t)|\calS|.
\]
Summing \eqref{eqn12} over all $j$, dividing both sides by $n$, and noting that
$|\calS|=S(n,t)$ (since $\calS$ was chosen to be minimal)
gives the desired inequality.
\end{proof}

\begin{corollary}
For all $0<t<n-1$, we have
\[
S(n,t)\geq \frac{1}{n-t-(t/n)+1/2} {n\choose t+1}.
\]
\end{corollary}

\begin{proof}
Simply use the facts that
\[
T(n-1,t+1,t)\geq \frac{1}{n-t-1} {n-1\choose t+1},
\]
and
\[
{n\choose t+1} = {n-1\choose t+1}+{n-1\choose t}.
\]
\end{proof}

Equivalently, if we let $k=n-t-1$, then we have that for all $0<k<n-1$,
\[
S(n,n-k-1)\geq \frac{1}{k+(k+1)/n+1/2} {n\choose k}.
\]

To relate this to the asymptotic results shown earlier, we note the following corollary.

\begin{corollary}
For all $k>0$, as $n\rightarrow\infty$, 
\[
S(n,n-k-1) \geq \bigl(1-o(1)\bigr) \frac{1}{k+1/2} {n\choose k}.
\]
\end{corollary}

\begin{proof}
Trivial.
\end{proof}

For fixed $k$, $T(n,n-k,n-k-1)$ is asymptotic to $\frac{1}{k+1}{n\choose k}$
(cf. \cite{stop:rodl85}).
So the above corollary shows that for \emph{any} fixed $k$, 
the ratio $S(n,n-k-1)\,/\,T(n,n-k,n-k-1)$ is bounded away from $1$
as $n\rightarrow\infty$.

\section{On the Schwartz-Vardy Conjecture}\label{sec3}

Schwartz and Vardy \cite{stop:schwartz06} conjectured
that the stopping redundancy of an MDS code only depends on its
length and minimum distance.

In \cite{stop:han07a}, we showed that for given $1<d\leq 5$ and $n\geq d$,
the leeway in $\rho(\C)$ for any $[n,n-d+1,d]$ MDS code $\C$ is at most one. 
We can now close the gap completely and show that the Schwartz-Vardy
conjecture is true for all MDS codes with $1<d\leq 5$, using the recurrent
bounds of Theorem~\ref{thm6}.

\begin{theorem}\label{thm12}
For all $n\geq 6$, 
\[
S(n,3)=T(n,4,3).
\]
\end{theorem}

\begin{proof}
Let $\T$ be a minimal Tur\'an $(n,4,3)$-system, $n\geq 6$.
We show that $\T$ must also be an $(n,3)$-SE system.
Since $\T$ is a Tur\'an $(n,4,3)$-system, all $4$-sets are covered. 
In \cite{stop:han07a}, it was shown that all $1$-sets and $2$-sets are covered too. 
It remains to show that all $3$-sets are covered. 
Suppose, to the contrary, that some $3$-set, $X$, is not covered. 
It was shown in \cite{stop:han07a} that at least $1+2{n-3\choose 2}$
blocks in $\T$ contain elements from $X$, and those that do not
form a Tur\'an $(n-3,4,3)$-system.
Hence, 
\[
S(n,3)\geq T(n,4,3)=|\T|\geq T(n-3,4,3)+2{n-3\choose 2} +1.
\]
However, from Theorem~\ref{thm6},
\[
S(n,3)\leq T(n-1,4,3)+T(n-2,3,2)+T(n-3,2,1)+1.
\]
And since (cf. \cite{stop:sidorenko97})
\[
T(n,s,t)\leq T(n-1,s,t)+T(n-1,s-1,t-1),
\]
we have
\[
T(n-1,4,3)\leq T(n-3,4,3)+T(n-3,3,2)+T(n-2,3,2).
\]
Together, these imply that
\[
S(n,3)\leq T(n-3,4,3)+2T(n-2,3,2)+T(n-3,3,2)+n-3.
\]
Putting the upper and lower bounds on $S(n,3)$ together, we have 
\[
2T(n-2,3,2)+T(n-3,3,2)+n-3\geq 2{n-3\choose 2} +1.
\]
However, since it is known (cf. \cite{stop:mantel07} \cite{stop:turan90})
that $T(n,3,2)=\lfloor n/2\rfloor(\lceil n/2\rceil -1)$,
the above inequality results in a contradiction for all $n\geq 9$. 
For $n=6,7,8$, the theorem can be verified directly, and was also proved 
in \cite{stop:han07a} (for $6\leq n\leq 53$) using a different argument.
\end{proof}

Summarizing Theorem~\ref{thm12} and results in \cite{stop:han07a},
we make the following conclusion.

\begin{proposition}
If $\C$ is an $[n,n-d+1,d]$ MDS code, and $1<d\leq 5$, then
\[
\rho(\C)=S(n,d-2).
\]
\end{proposition}

\begin{proof}
For $(n,d)\in\{(4,4),(5,4),(5,5)\}$, it is easy to find that
$S(4,2)=3$, $S(5,2)=5$, $S(5,3)=4$, and verify that $\rho(\C)=S(n,d-2)$
in all three cases.
Otherwise, since $T(n,d-1,d-2)\leq \rho(\C)\leq S(n,d-2)$,
it follows immediately from 
Theorem~\ref{thm12} and \cite[Theorem~14, Theorem~16]{stop:han07a} that
\[
\rho(\C)=S(n,d-2)=T(n,d-1,d-2)
\]
in all other cases.
\end{proof}

In addition, from Corollary~\ref{cor1} and Corollary~\ref{cor2},
we see that as $n\rightarrow\infty$, if $d=o(\sqrt{n})$,
or if $k=o(\sqrt{n})\rightarrow\infty$, then
\[
\rho(\C)\sim S(n,d-2).
\]
So in these cases we may say that the Schwartz-Vardy conjecture holds
in the asymptotic sense.

Our approach regarding the conjecture has been so far to show that in some cases
the upper and lower bounds on stopping redundancy
(SE and Tur\'an numbers, respectively) converge,
either exactly or asymptotically.
However, we have seen that in other cases the corresponding
SE and Tur\'an numbers can be
provably different, even in the asymptotic sense
(for example, when $k$ is a fixed constant),
which shows the limitation of the current approach 
in fully resolving the conjecture. 

In fact, it is our belief (cf. \cite{stop:han06}) that 
for an $[n,n-d+1,d]$ MDS code $\C$, 
\[
\rho(\C)=S(n,d-2),
\]
the proof of which would in turn prove the Schwartz-Vardy conjecture.
We have shown that this is true (or close to be true) when either $d$ or $k$
is $o(\sqrt{n})$.
A reasonable question to ask is: what if both $d$ and $k$ are larger than $o(\sqrt{n})$?
For example, what if $k/n$ approaches a constant?
The current approach only bounds $\rho(\C)$ to within a factor of up to  $\ln n$.
For example, using the result of Theorem~\ref{thm7}, we have
\[
\rho(\C)
\leq \frac{n-d+2}{n-d+1} \cdot \frac{1+\ln (n-d+1)}{d-1} \cdot {n\choose d-2},
\]
while, for the lower bound, we saw that 
\[
\rho(\C)\geq \frac{1}{d-1}{n\choose d-2}.
\]

Alternatively, let's make the following observation.
Suppose we are given one optimal parity-check matrix, i.e. one with
$\rho(\C)$ rows that maximizes stopping distance.
It is not apparent that all rows should have minimum weight, but suppose
$T'$ rows are of minimum weight and the rest are not.
We can replace each row that is not of minimum weight (and whose weight is,
of course, at most $n$) with no more than $\lceil n/(n-d+2)\rceil$ 
minimum-weight rows, such that the union of supports of these rows is precisely the support of the row they replaced.
It is simple to verify that the replacement procedure does not decrease
the stopping distance, which also implies that the rank of the matrix is not reduced.
After all rows that are not of minimum weight have been replaced, we obtain a parity-check matrix with at most 
\[
T'+\left\lceil \frac{n}{n-d+2}\right\rceil \bigl(\rho(\C)-T'\bigr)
\]
rows, all having minimum weight, that achieves maximum stopping distance. 
Therefore, 
\[
T'+\left\lceil \frac{n}{n-d+2}\right\rceil \bigl(\rho(\C)-T'\bigr) \geq S(n,d-2).
\]
Now note that
\[
T'\geq T(n,d-1,d-2),
\]
so we have
\[
\rho(\C)\geq \frac{S(n,d-2)}{\bigl\lceil \frac{n}{n-d+2}\bigr\rceil}
  + \frac{\bigl\lceil \frac{n}{n-d+2}\bigr\rceil-1}{\bigl\lceil \frac{n}{n-d+2}\bigr\rceil}
     \cdot T(n,d-1,d-2).
\]
Without knowing better how $T(n,d-1,d-2)$ compares with $S(n,d-2)$,
if we just ignore the second term, we obtain
\[
\frac{S(n,d-2)}{\bigl\lceil \frac{n}{n-d+2}\bigr\rceil}
\leq \rho(\C)
\leq S(n,d-2).
\]
This shows that in many cases $S(n,d-2)$ is a good estimate of $\rho(\C)$.
For example, if the code rate $R=(n-d+1)/n\geq 1/2$, then
\[
\frac{1}{2} S(n,d-2) \leq \rho(\C) \leq S(n,d-2).
\]
And, clearly, for any constant code rate, $\rho(\C)$ is within a constant factor
of $S(n,d-2)$.

\section{Concluding Remarks}\label{sec4}

While we have obtained a fairly good understanding of the SE number,
and in some cases come close to uncovering the true value of the stopping redundancy
of MDS codes along the way, many interesting questions remain unanswered. 
For example, what is the asymptotic value of $S(n,n-k-1)$ for a fixed $k$?
(Is it asymptotic to $\frac{1}{k}{n\choose k}$?)
And how does $S(n,t)$ compare with $T(n,t+1,t)$ in general?
(Do they differ by at most a constant factor?)
Finally, is it true that the stopping redundancy of an $[n,n-d+1,d]$ MDS code
equals $S(n,d-2)$ (as we conjectured in \cite{stop:han06})?

\useRomanappendicesfalse
\appendices

\section{Asymptotics of \eqref{eqn7} for~~$t<n-\ln n$}
\label{app1}

Rewrite \eqref{eqn7} as
\begin{equation}\label{eqn14}
\eta(n,t)
= \frac{n-t}{t+1}\cdot \sum^{t+1}_{i=1} f(i)
\end{equation}
where
\[
f(i)=\frac{{t+1\choose i}}{{n-i\choose n-t-1}} \cdot
       (n-t)^{-\frac{i}{t}{n-i\choose n-t-1}}
\]
Assuming
\[
n-t > \ln n,
\]
we show that the $(t+1)$-st term, 
\[
f(t+1)=(n-t)^{-(t+1)/t},
\]
prevails in the sum of \eqref{eqn14}.

It suffices to show that each of the other $t$ terms is a $o(1/t)$ fraction of $f(t+1)$.
For $i=t$, it is easy to verify that
\[
f(t)=\frac{t+1}{n-t} \cdot (n-t)^{-(n-t)}=o(1/t)f(t+1).
\]
In general, we have
\begin{equation}\label{eqn15}
\frac{f(i)}{f(t+1)}=
\frac{{t+1\choose i}}{{n-i\choose n-t-1}}\cdot
(n-t)^{-\frac{i}{t}{n-i\choose n-t-1}+\frac{t+1}{t}}.
\end{equation}
For $i=1,\ldots,t-1$, consider two cases.
For $i\leq 2n/\ln n=o(n)$, we have
\[
{n-i\choose n-t-1} \geq {n(1-2/\ln n)\choose 2} = \Omega(n^2).
\]
In this case, \eqref{eqn15} decreases super-exponentially with $n$, 
and is certainly $o(1/t)$.
On the other hand, for $2n/\ln n <i\leq t-1$, 
we show that $f(i)<f(t)=o(1/t)f(t+1)$,
by showing that $f(i)$ is monotonically increasing for
$2n/\ln n <i\leq t$.
Indeed, for $2n/\ln n <i\leq t-1$, we have
\begin{align*}
\frac{f(i+1)}{f(i)}&= \frac{n-i}{i+1}\cdot
(n-t)^{-\frac{i+1}{t}{n-i-1\choose n-t-1}+\frac{i}{t}{n-i\choose n-t-1}}\\
&= \frac{n-i}{i+1}\cdot
(n-t)^{\frac{1}{t}\left(i-\frac{t-i+1}{n-t-1}\right){n-i-1\choose n-t-2}}\\
&> \frac{\ln n}{n}\cdot
(\ln n)^{\frac{1}{t}\left(\frac{2n}{\ln n}-\frac{n}{\ln n}\right){\ln n\choose 2}}\\
&> \left(\frac{\sqrt{\ln n}}{e}\right)^{\ln n}\\
&> 1.
\end{align*}


\section{Asymptotics of the Bound \eqref{eqn3} for~~$t\geq n-\ln n$}
\label{app2}

Let $k=n-t-1$, and $j=t+1-i$.
The right-hand side of \eqref{eqn3} becomes
\begin{equation}\label{eqn8}
p{n\choose k+1}+
\sum^{t}_{j=0} {n\choose k+j} (1-p)^{(n-k-j){k+j\choose k}}.
\end{equation}
Let $p=(\ln n)/n$.
Assume that
\[
1<k=o\left(\frac{n\ln\ln n}{\ln n}\right)
\]
($n-\ln n\leq t <n-2$ being a special case).
We show that the first term in \eqref{eqn8} prevails.
Note that
\begin{align*}
\left(1-\frac{\ln n}{n}\right)^{(n-k-j){k+j\choose k}}
&\leq e^{-\frac{\ln n}{n}\cdot (n-k-j){k+j\choose k}}\\
&= n^{-\frac{n-k-j}{n}{k+j\choose k}}.
\end{align*}
For $1\leq j\leq n\sqrt{2/\ln n}$, we have
\[
\frac{n-k-j}{n} {k+j\choose k}
\geq \bigl(1-o(1)\bigr) {j+2\choose 2}
> j+1.
\]
For $j> n\sqrt{2/\ln n}$, we have
\[
\frac{\ln n}{n}\cdot (n-k-j) {k+j\choose k}
\geq \frac{\ln n}{n} {j+2\choose 2}
> n.
\]
Noting further that for $j\geq 1$, 
$$
{n\choose k+j}\leq {n\choose k+1}\cdot n^{j-1}
$$
we see that the second term in \eqref{eqn8} is at most
\begin{align*}
& {n\choose k} \cdot n^{-\frac{n-k}{n}} +
   \sum^{\left\lfloor n\sqrt{\frac{2}{\ln n}}\right\rfloor}_{j=1} 
   {n\choose k+1}\cdot n^{j-1} \cdot n^{-\frac{n-k-j}{n}{k+j\choose k}}\\
 &+\sum^{r}_{j=\left\lfloor n\sqrt{\frac{2}{\ln n}}\right\rfloor+1}
                 {n\choose k+j}\cdot e^{-\frac{\ln n}{n}\cdot (n-k-j){k+j\choose k}}\\
&\leq {n\choose k+1}\cdot n^{-1+\frac{k}{n}} +
   {n\choose k+1}\sum^{\left\lfloor n\sqrt{\frac{2}{\ln n}}\right\rfloor}_{j=1} n^{-2}\\
   &\quad\ + e^{-n}\sum^{r}_{j=\left\lfloor n\sqrt{\frac{2}{\ln n}}\right\rfloor+1} 
      {n\choose k+j}\\
&\leq \frac{1}{n} {n\choose k+1} \left(e^{\frac{k\ln n}{n}}+
   \sqrt{\frac{2}{\ln n}}
   \right)
   + 2^n\cdot e^{-n}\\
&= \frac{1}{n} {n\choose k+1} \Bigl(o(\ln n)+o(1)\Bigr) + o(1)\\
&= o\left(\frac{\ln n}{n} {n\choose k+1}\right).
\end{align*}
Hence, at $p=(\ln n)/n$, the upper bound \eqref{eqn3}
is asymptotic to $\frac{\ln n}{n} {n\choose t}$ for all $t$ such that
$2<n-t=o\bigl((n\ln\ln n)/\ln n\bigr)$,
which implies that for $2<n-t=o\bigl((n\ln\ln n)/\ln n\bigr)$, the upper bound \eqref{eqn3},
when minimized over $p$, is $O\bigl(\frac{\ln n}{n} {n\choose t}\bigr)$.

Note that the $O\bigl(\frac{\ln n}{n} {n\choose t}\bigr)$ estimate is not always tight.
For example, when $t\approx n-\ln n$, we have shown using a different analysis
that the upper bound \eqref{eqn3},
when minimized over $p$, is in fact $O\bigl(\frac{\ln\ln n}{n} {n\choose t}\bigr)$.
However, note that by keeping just the $i=t$ term in the sum, 
the upper bound \eqref{eqn3} is at least
\begin{align*}
&\left(p+(1-p)^{t(n-t)}\right) {n\choose t}\\
&\geq \left(1-\bigl(1-1/a\bigr)\cdot a^{-1/(a-1)}\right){n\choose t}\\
&= \left[\frac{\ln a}{a}+O\left(\frac{1}{a}\right)\right] {n\choose t},
\end{align*}
where $a=t(n-t)$.
In particular, this shows that the $O\bigl(\frac{\ln n}{n} {n\choose t}\bigr)$
estimate is tight if $n-t$ is $\Theta(1)$.
That is, the bound \eqref{eqn3} is
$\Theta\bigl(\frac{\ln n}{n}{n\choose t}\bigr)$ when minimized over $p$,
for all $t=n-\Theta(1)$.

\section*{Acknowledgment}
This work was supported in part by the Center for Magnetic Recording
Research at UCSD, and by grant No.~2002197 from
the United States--Israel Binational Science Foundation (BSF),
Jerusalem, Israel. The authors wish to thank Tuvi Etzion for
helpful discussions.

\bibliographystyle{IEEEtran}
\bibliography{IEEEabrv,math,stop}

\end{document}